\documentclass[floatfix,aps,twocolumn,prb,amsmath,amssymb,showpacs,superscriptaddress]{revtex4-1}

\usepackage{eurosym}
\usepackage{amssymb}
\usepackage{amsmath}
\usepackage{amsfonts}
\usepackage{amsbsy}
\usepackage{graphicx}
\usepackage{bm}
\usepackage{color}
\usepackage{hyperref}
\usepackage{units}

\begin{document}

\title{Low-energy physics for an iron phthalocyanine molecule on Au(111)}

\author{A. A. Aligia}
\affiliation{Instituto de Nanociencia y Nanotecnolog\'{\i}a CNEA-CONICET,
Centro At\'{o}mico Bariloche and Instituto Balseiro, 8400 Bariloche, Argentina}

\begin{abstract}
The system of an iron phthalocyanine molecule on the Au(111) surface,
has been studied recently due to its peculiar properties. In particular, 
several surprising results of scanning tunneling spectroscopy changing
the position of the molecule and applying magnetic field can be explained
by the {\it non-Landau} Fermi liquid state of a 2-channel spin-1 Kondo model with anisotropy. 
The localized orbitals near the Fermi level are three, one of symmetry 
$z^2$ and two (nearly) degenerate $\pi$ orbitals of symmetry 
$xz$ and $yz$. Previous studies using the numerical 
renormalization group neglected one of these orbitals to render
the problem tractable. Here we investigate, using a slave-boson mean-field 
approximation, if the splitting $S$ between $\pi$ orbitals 
caused by spin-orbit coupling (SOC) justifies this approximation.
We obtain an abrupt transition from a 3-band regime to a 2-band one
at a value of $S$ which is about 1/3 of the atomic SOC for Fe,
justifying the 2-band model for the system.

\end{abstract}


\maketitle


\section{Introduction}
\label{intro}

The Kondo effect, found first in metals containing magnetic impurities~\cite{hewson97,kondo64}, is a paradigmatic example of a strongly correlated system in condensed 
matter physics. In the simplest form, it arises when the free electrons of a metallic host
screen completely the magnetic moment of an impurity (under- and over-screening are not discussed here~\cite{nozieres80}).

The high resolution and atomic control of the scanning tunneling microscope (STM) allow experimentalists to deposit magnetic molecules on metallic surfaces leading to a large class of realizations of Kondo phenomena 
in which the current can be controlled by different 
external parameters \cite{liang02,parks07,roch08,oso2,parks10,florens11,
vincent12,mina,orma,giro,hira,yang,moro,granet,guo}. This subject is of interest for its potential  use in new electronic devices \cite{naturefocus,evers20}.
The  differential conductance $G(V)=dI/dV$ as a function of the sample bias $V$, where $I$ is the current flowing
through the STM provides information
of the low-energy electronic structure of the system. This technique is called scanning tunneling spectroscopy.

This spectroscopy for FePc on Au(111) at low temperature shows several striking 
features. $G(V)$ around $V=0$ shows a broad peak of half width $\sim 20$ meV
and mounted on it a dip nearly two orders of magnitude narrower \cite{mina,hira,yang}. Application of a magnetic field $B$ transforms the dip
into a peak \cite{yang}. A similar behavior is observed for MnPc on Au(111)\cite{guo}. When the molecule is raised from the surface, weakening the Kondo effect, the dip broadens \cite{hira}. All these features have been recently
explained in a consistent fashion by a 2-channel spin-1 Kondo model with anisotropy \cite{nl}. This is an extension to non-degenerate channels of a model for Ni impurities in a Au chain \cite{blesio18,blesio19}. 

For $B=0$, these models display a topological quantum phase transition  between an ordinary Fermi liquid with a peak in the spectral density 
at the Fermi energy and a
{\it non-Landau} Fermi liquid with a pseudogap at the Fermi level, topologically characterized by a non-trivial Friedel sum rule with non-zero Luttinger integrals. The results of Ref. \onlinecite{nl} indicate that 
FePc on Au(111) is near the topological transition at the {\it non-Landau} side of it. Pressing the molecule against the substrate by the STM tip should induce the transition.

An abrupt transition from a peak to a dip in the spectral density has been found previously
in other two-orbital \cite{deleo} and two-impurity \cite{curtin18,nishi} Anderson models.
In the latter, the transition was also ascribed as due to a jump 
in Luttinger integrals \cite{curtin18,nishi}. 

LDA+U calculations \cite{mina} indicate that the configuration of Fe in the system is $(d_{xy})^2(d_{z^2})^1)(d_{\pi})^3$. Therefore the partially filled
orbitals can be described as one hole with $z^2$ symmetry and one hole with 
$\pi$ ($xz$ or $yz$) symmetry. They are coupled forming a spin $S=1$ by the Hund rules. If this configuration is hybridized with the different excited configurations with one hole, one has a 3-channel Anderson model, which 
has been studied by Fern\'andez \textit{et al. } \cite{joaq} using a 
slave-boson mean-field approximation (SBMFA). 
This model is justified in more detail in Sec. II of this work.
In the limit of small
hybridization compared with the difference between the energies of both
configurations, the model is equivalent to a 3-channel $S=1$ Kondo model,
which is more involved than the 2-channel Kondo model used to describe
the system as a {\it non-Landau} Fermi liquid  \cite{nl}. 
The 2-channel Kondo model has been chosen because the 3-channel case 
is very difficult to treat by the numerical 
renormalization group due to the huge increase in the Hilbert space
at each iteration, while the essence of the physics is expected
to be captured by the 2-channel model \cite{nl}.

However, Fe has 
a spin-orbit coupling (SOC) $\sim 76$ meV \cite{socfe} disregarded before.
In first order, the SOC splits the $\pi$ states by this amount and leaves 
the $z^2$ orbitals unchanged. This splitting might justify the 2-channel model,
supporting the quantitative validity of previous results. Naively,
one would think that comparing the magnitude of the SOC with another energy
scale might solve the issue. However, this is not so simple. Note that
for degenerate channels in the 2-channel spin-1 Kondo model with easy plane anisotropy $DS_{z}^{2}$, 
the topological quantum phase transition takes place for $D=D_c \sim 2.5 T_K$, where
$T_K$ is the Kondo temperature \cite{blesio18}. However for inequivalent
channels, there is no simple relation between the critical anisotropy 
$D_c$ and the Kondo temperatures for both channels \cite{nl}.

In this work we study the 3-channel Anderson model including a splitting 
$S$ between the $\pi$ orbitals. Due the complexity of the problem, we treat
it in a SBMFA described below. Since the approximation is unable to describe
the {\it non-Landau} Fermi liquid, we neglect the anisotropy $D$. 
We modify previopus treatments of the SMBFA in such a way that the correct 
Kondo temperature is reproduced in known limits. 
We note that alternative methods, like the non-crossing approximation
for this two-channel case (see appendix of Ref. \onlinecite{blesio19})
or equations of motion \cite{roer} do not reproduce the correct Kondo temperature.

The results indicate that 
the excited $\pi$ orbital can be neglected for $S>S_c$, where the critical
value $S_c$ is about a third of the SOC for Fe, 
supporting the validity of the 2-channel model.

This work is organized as follows. In section \ref{model} we describe the
3-band Anderson model used for this study and justify it in detail. 
In section \ref{apro} we explain
the SBMFA used to solve the model in the Kondo or integer valence limit. The
results are presented in Section \ref{res} and  Section \ref{sum} contains a summary and a discussion.

\section{Model}

\label{model}

We describe the system by an Anderson model containing two magnetic
configurations. This implies that we take the limit of infinite Coulomb repulsion.
This approximation does not affect the essential physics at low energies, 
including the Kondo effect. The neglected configurations might affect slightly
the parameters of the model (discussed at the beginning of Section \ref{res}),
but in any case they are adjusted form low-energy experimental results.

The ground state corresponds essentially to the $3d^{6}$
configuration of Fe with one hole in the 3$d_{z^{2}}$ orbital and another
hole in a $\pi $ orbital (3$d_{xz}$ or 3$d_{yz}$) forming a triplet. 
The $d_{xy}$ orbitals are occupied by both spins and the $d_{x^{2}-y^{2}}$ are empty. It is known that to have a Kondo effect, one needs either spin
degeneracy or orbital degeneracy in the ground-sate configuration of the
magnetic "impurity" (the FePc molecule in our case). Therefore, the
completely filled $d_{xy}$ orbitals and the empty $d_{x^{2}-y^{2}}$ orbitals 
might affect slightly the parameters of the low-energy effective model, 
but not the form of it. 

The three relevant 3d orbitals of Fe  have some admixture with linear
combinations of orbitals of neighboring N atoms,  forming molecular states
of the same symmetry \cite{lobos}. In turn, these states also hybridize with surface and conduction bands of the same symmetry. In particular the surface band can be accurately described by a free electron band with plane waves $\exp \left[ i\left( k_{x}x+k_{y}y\right) \right] $ \cite{hul,limot,agam,fermo}.
Neglecting the discrete nature of the substrate, the symmetry of the system is $C_{4v}$.
The molecular orbital with symmetry $z^{2}$ belongs to the irreducible
representation $A_{1}$ of $C_{4v}$ and hybridizes only with surface states
(combination of plane waves) of the form 
$\cos \left( k_{x}x\right) \cos \left( k_{y}y\right) $ which also belong to $A_{1}$. 
Similarly the
molecular orbital with $xz$ symmetry (a component of the two-dimensional
irreducible representation $E$) hybridizes only with surface states of the
form $\sin \left( k_{x}x\right) \cos \left( k_{y}y\right) $, and for the  $xy$ symmetry the corresponding conduction states are 
$\cos \left(k_{x}x\right) \sin \left( k_{y}y\right) $. 
These arguments can be extended to the bulk conduction states.  

The above arguments justify the 3-channel model studied previously by 
Fern\'{a}ndez \textit{et al. } \cite{joaq}. 
However, here assume that the $\pi $ orbitals are split by an energy $S$. 
We call $a$ ($b$) the linear combination of 
$\pi $ holes with lower (larger) energy. The other configuration, the $3d^{7}$ one, has a hole in either the 3$d_{z^{2}}$ orbital or in a $\pi $ orbital. 
We denote the
two spin triplets by $|aM\rangle $ and $|bM\rangle $, depending on which 
$\pi $ orbital is occupied by a hole, in addition to the $z^2$ one, where $M$ is the spin-1 projection.
Similarly, the three spin doublets are represented by 
$|z^{2}\sigma \rangle $, $|a\sigma \rangle $, and  $|b\sigma \rangle $, where $\sigma $ is the
spin-1/2 projection. Both configurations are mixed via hybridization with
the conduction bands. The model is an extension to finite $S$ of that
considered previously by Fern\'{a}ndez \textit{et al. } \cite{joaq}

The Hamiltonian is 
\begin{widetext}
\begin{eqnarray}
H &=&H_{\mathrm{mol}}+H_{\mathrm{band}}+H_{\mathrm{mix}},  \notag \\
H_{\mathrm{mol}} &=&E_{z}\sum_{\sigma }|z^{2}\sigma \rangle \langle
z^{2}\sigma |+\sum_{\pi \sigma }E_{\pi }|\pi \sigma \rangle \langle \pi
\sigma |+\sum_{\pi M}(E_{\pi }+E_{d})|\pi M\rangle \langle \pi M|  \notag \\
H_{\mathrm{band}} &=&\sum_{k\nu \sigma }\epsilon _{k\nu \sigma }c_{k\nu
\sigma }^{\dagger }c_{k\nu \sigma }  \notag \\
H_{\mathrm{mix}} &=&\sum_{\pi k}\sum_{\sigma \sigma ^{\prime }M}t_{\pi
}\left\langle \frac{1}{2}\frac{1}{2}\sigma \sigma ^{\prime }|1M\right\rangle
\left( c_{k\pi \sigma }^{\dagger }|z^{2}\sigma ^{\prime }\rangle \langle \pi
M|+\mathrm{H.c.}\right)   \notag \\
&-&\sum_{\pi k}\sum_{\sigma \sigma ^{\prime }M}t_{z}\left\langle 
\frac{1}{2}\frac{1}{2}\sigma \sigma ^{\prime }|1M\right\rangle \left( c_{kz\sigma
}^{\dagger }|\pi \sigma ^{\prime }\rangle \langle \pi M|
+\mathrm{H.c.} \right) ,  \label{ham}
\end{eqnarray}
\end{widetext}
where $H_{\mathrm{mol}}$ represent the molecular states, with $E_{\pi }=E_{a}
$ for $\pi =a$, $E_{b}=E_{a}+S$, and $E_{d}<0$ is the difference between the
energies of the lowest lying states of both configurations.  $H_{\mathrm{band}}$ represents the three conduction bands, with the same symmetry as the
corresponding molecular states $(\nu =z^{2}$, $a$ or $b$). $H_{\mathrm{mix}}$
describes the mixing Hamiltonian (also called hybridization) 
in terms of Clebsch-Gordan coefficients and
two hopping amplitudes (we assume $t_{a}=t_{b}$). 

In general, the origin of the splitting $S$ could be either a symmetry
breaking which renders the orbital with symmetries $xz$ and $yz$
inequivalent or SOC or both. If the origin is the SOC, the states $|\pi
\sigma \rangle $with one hole in the $\pi $ orbitals are (except for an
irrelevant phase)

\begin{eqnarray}
|a &\uparrow &\rangle =\frac{|xz\uparrow \rangle +i|yz\uparrow \rangle }
{\sqrt{2}},|a\downarrow \rangle =\frac{|xz\downarrow \rangle -i|yz\downarrow
\rangle }{\sqrt{2}},  \notag \\
|b &\uparrow &\rangle =\frac{|xz\uparrow \rangle -i|yz\uparrow \rangle }
{\sqrt{2}},|b\downarrow \rangle =\frac{|xz\downarrow \rangle +i|yz\downarrow
\rangle }{\sqrt{2}},  \label{states}
\end{eqnarray}
and similarly for the triplet states  $|\pi M\rangle $ combining with a $z^{2}
$ hole to build a spin triplet.

In the limit in which only one multiplet is relevant for each configuration
(very large $S$ and $|E_{z}-E_{a}|$) the model has been solved exactly by
the Bethe ansatz \cite{bethe}. We use this result to refine the SBMFA. 

\section{Slave bosons in mean-field approximation (SBMFA)}

\label{apro} 
We solve the model using a slave-boson treatment similar to
that of Kotliar and Ruckenstein (KR) \cite{kr} in the mean-field approximation. 
This treatment has severe limitations when a magnetic field in an arbitary 
direction is applied or when finite Coulomb interactions in the multiorbital
case are considered. In these cases the rotationally invariant slave-boson formalism
is more convenient \cite{lecher}. The disadvantage of this method is that it 
introduces more bosonic variables and determining them minimizing the energy 
becomes more involved. Fortunately in our case in which infinite Coulomb
repulsion is implicitly assumed and no magnetic field is applied, the KR formalism 
can  be applied.

The KR approach consists of introducing bosonic operators for each of the
states in the fermionic description. In this representation, in our case, 
we can write
the doublets using bosons $s_{\nu \sigma }^{\dagger }$ which correspond to
the singly occupied states

\begin{eqnarray}
|\pi \sigma \rangle  &\leftrightarrow &f_{\pi \sigma }^{\dagger }s_{\pi
\sigma }^{\dagger }|0\rangle   \notag \\
|z^{2}\sigma \rangle  &\leftrightarrow &f_{z\sigma }^{\dagger }s_{z\sigma
}^{\dagger }|0\rangle ,  \label{si}
\end{eqnarray}
where $f_{\pi \sigma }^{\dagger }$ ($f_{z\sigma }^{\dagger }$) is a fermionic
operator that creates a localized hole with $\pi $ ($z^{2}$) symmetry. The
triplets are represented using bosons $d_{\pi M}^{\dagger }$ for the doubly
occupied states as follows

\begin{eqnarray}
|\pi 1\rangle  &\leftrightarrow &d_{\pi 1}^{\dagger }f_{\pi \uparrow
}^{\dagger }f_{z\uparrow }^{\dagger }|0\rangle   \notag \\
|\pi 0\rangle  &\leftrightarrow &\frac{1}{\sqrt{2}}d_{\pi 0}^{\dagger
}\left( f_{\pi \uparrow }^{\dagger }f_{z\downarrow }^{\dagger }+f_{\pi
\downarrow }^{\dagger }f_{z\uparrow }^{\dagger }\right) |0\rangle   \notag \\
|\pi -1\rangle  &\leftrightarrow &d_{\pi -1}^{\dagger }f_{\pi \downarrow
}^{\dagger }f_{z\downarrow }^{\dagger }|0\rangle ,  \label{do}
\end{eqnarray}

The Hamiltonian in this representation takes the form

\begin{eqnarray}
H &=&E_{z}\sum_{\sigma }s_{z\sigma }^{\dagger }s_{z\sigma }+\sum_{\pi \sigma
}E_{\pi }s_{\pi \sigma }^{\dagger }s_{\pi \sigma }  \notag \\
&&+\sum_{\pi M}(E_{\pi }+E_{d})d_{\pi M}^{\dagger }d_{\pi M}+\sum_{k\nu
\sigma }\epsilon _{k\nu \sigma }c_{k\nu \sigma }^{\dagger }c_{k\nu \sigma } 
\notag \\
&&+(t_{\pi }\sum_{\pi \sigma }\left[ f_{\pi \sigma }^{\dagger }c_{\pi \sigma
}\left( d_{\pi 2\sigma }^{\dagger }s_{z\sigma }+\frac{1}{\sqrt{2}}d_{\pi
0}^{\dagger }s_{z\bar{\sigma}}\right) O_{\pi }\right]   \notag \\
&&+t_{z}\sum_{\pi \sigma }\left[ f_{z\sigma }^{\dagger }c_{z\sigma }\left(
d_{\pi 2\sigma }^{\dagger }s_{\pi \sigma }+\frac{1}{\sqrt{2}}d_{\pi
0}^{\dagger }s_{\pi \bar{\sigma}}\right) O_{z}\right]   \notag \\
&&+\mathrm{H.c.}),  \label{hsb}
\end{eqnarray}
where the operators $O_{\nu }=1$ in the physical subspace (they are defined
below) and the following constraints should be satisfied to restrict the
bosonic Hilbert space to the physical subspace

\begin{eqnarray}
1 &=&\sum_{\sigma }\left( \sum_{\pi }s_{\pi \sigma }^{\dagger }s_{\pi \sigma
}+s_{z\sigma }^{\dagger }s_{z\sigma }\right) +\sum_{\pi M}d_{\pi M}^{\dagger
}d_{\pi M},  \notag \\
f_{\pi \sigma }^{\dagger }f_{\pi \sigma } &=&s_{\pi \sigma }^{\dagger
}s_{\pi \sigma }+d_{\pi 2\sigma }^{\dagger }d_{\pi 2\sigma }
+\frac{1}{2}d_{\pi 0}^{\dagger }d_{\pi 0},  \notag \\
f_{z\sigma }^{\dagger }f_{z\sigma } &=&s_{z\sigma }^{\dagger }s_{z\sigma
}+\sum_{\pi }\left( d_{\pi 2\sigma }^{\dagger }d_{\pi 2\sigma }
+\frac{1}{2}d_{\pi 0}^{\dagger }d_{\pi 0}\right) .  \label{cons}
\end{eqnarray}

The idea of the introduction of the operators $O_{\nu }$ is to correct the
mean-field solution so that certain limits are reproduced. For the Hubbard
model, Kotliar and Ruckenstein have chosen the corresponding operators in
such a way that the non-interacting limit is reproduced. In this case,
the approximation becomes equivalent to the Gutzwiller approximation \cite{kr}.
However, this choice, even in the one-channel case, leads to a too large Kondo 
temperature for large Coulomb repulsion \cite{notesb}. In our model, 
this repulsion is infinite since only two neighboring configurations 
of the localized states are retained. Therefore, we determine the $O_{\nu }$
requiring that when only one hybridization channel is relevant and in
the Kondo limit (small relevant $t_{\nu }$ compared to the difference
between the smallest energies of both configurations), 
the correct exponent of the Bethe ansatz result \cite{bethe} for the corresponding Kondo temperature is reproduced

\begin{eqnarray}
T_{K}^{\pi } &\sim &\Delta _{\nu }\exp \left[ \frac{\pi \left( E_{d}+E_{\pi
}-E_{z}\right) }{2\Delta _{\pi }}\right] ,\text{ }  \notag \\
T_{K}^{z} &\sim &\Delta _{z}\exp \left[ \frac{\pi E_{d}}{2\Delta _{z}}\right]
,  \label{tkb}
\end{eqnarray}
where $\Delta _{\nu }=\pi \rho _{\nu }t_{\nu }^{2}$, with $\rho _{\nu }$ the
density of conduction electrons with symmetry $\nu $, 
is called the resonant-level width for orbitals of symmetry $\nu$
and concides with half of the width at half maximum of
the corresponding peak in the spectral density of the molecular
orbitals with symmetry $\nu$ in the non-interacting case. 
In the SBMFA results the prefactor is replaced by the half band width $D$, 
but this is not essential and the important point is to recover the correct 
exponent.

In addition we ask
that when the $\pi $ states are degenerate ($S=0$) and $t_{z}=0$ (2-channel
degenerate case) the exponent in  $T_{K}^{\pi }$ is doubled (generalizing
the SU(4) case \cite{joaq}). These limiting cases can be easily treated as
in Ref. \onlinecite{joaq}. We find that a possible choice is

\begin{eqnarray}
O_{\pi } &=&[1-A\sum_{\pi M}d_{\pi M}^{\dagger }d_{\pi M}  \notag \\
&&-B\left( \sum_{M}d_{aM}^{\dagger }d_{aM}\right) \left(
\sum_{M}d_{bM}^{\dagger }d_{bM}\right) ]^{-1/2},  \notag \\
O_{z} &=&\left( 1-A\sum_{\pi M}d_{\pi M}^{\dagger }d_{\pi M}\right) ^{-1/2}.,
\label{onu}
\end{eqnarray}
with $B=2(1+1/\sqrt{2})^{2}/3\simeq 1.9428$, $A=1-B/2\simeq 0.0286$. For
simplicity and without affecting the semiquantitative validity of our
results, we take $B=2$, $A=0$, implying $O_{z}=1$. 

In the mean-field approximation, the bosonic operators are replaced by
numbers. Since we consider magnetic field $B=0$, these numbers do not depend
on spin projection. Then, there are five independent bosonic 
variables ($s_{z}$, $s_{a}$, $s_{b}$, $d_{a}$, $d_{b}$). Using the first constraint Eq. (\ref{cons}) we eliminate $s_{z}$ [see last Eq. (\ref{delef})]  keeping the
formalism symmetric in the $\pi $ ($a$ or $b$) variables.  The remaining
constraints are treated introducing three Lagrange multipliers $\lambda
_{\nu }$ and adding to the Hamiltonian the term 

\begin{eqnarray*}
H_{\mathrm{cont}} &=&\lambda _{z}\sum_{\sigma }\left( f_{z\sigma }^{\dagger
}f_{z\sigma }-\frac{1}{2}+\sum_{\pi }s_{\pi }^{2}\right)  \\
&&+\sum_{\pi \sigma }\lambda _{\pi }\left( f_{\pi \sigma }^{\dagger }f_{\pi
\sigma }-s_{\pi }^{2}-\frac{3}{2}d_{\pi }^{2}\right) .
\end{eqnarray*}

The problem is reduced to a non-interacting fermionic Hamiltonian, where the
seven variables $s_{\pi }$, $d_{\pi }$ and $\lambda _{\nu }$ are obtained
minimizing the energy (we take zero temperature). Assuming constant density
of conduction states $\rho _{\nu }$ extending from $-D$ to $D$, where the
Fermi energy lies at zero, the Green functions of the pseudofermions take a
simple form

\begin{equation}
G_{fv\sigma }(\omega )=\langle \langle f_{\nu \sigma };f_{\nu \sigma
}^{\dagger }\rangle \rangle =\frac{1}{\omega -\lambda _{\nu }+i\tilde{\Delta}_{\nu }},  \label{gnu}
\end{equation}
where the renormalized half width of the resonances $\tilde{\Delta}_{\nu}$ 
(which determine the
three Kondo scales) are 

\begin{eqnarray}
\tilde{\Delta}_{z} &=&\Delta _{z}\left( 1+\frac{1}{\sqrt{2}}\right)
^{2}\left( \sum_{\pi }s_{\pi }d_{\pi }\right) ^{2},  \notag \\
\tilde{\Delta}_{\pi } &=&\Delta _{\pi }\left( 1+\frac{1}{\sqrt{2}}\right)
^{2}\frac{d_{\pi }^{2}s_{z}^{2}}{1-18d_{a}^{2}d_{b}^{2}},  \notag \\
\text{with }s_{z}^{2} &=& \frac{1-\sum_{\pi }(2s_{\pi }^{2}
+3d_{\pi }^{2})}{2}.  \label{delef}
\end{eqnarray}

Using these Green functions, the change in energy after adding the impurity
can be evaluated easily as in similar problems using the SBMFA 
\cite{hewson97,joaq}. The result is

\begin{widetext}
\begin{eqnarray}
\Delta E &=&E_{z}-\lambda _{z}+2\sum_{\pi }\left( E_{\pi }-E_{z}+\lambda
_{z}-\lambda _{\pi }\right) s_{\pi }^{2}+3\sum_{\pi }\left( E_{\pi
}+E_{d}-\lambda _{\pi }\right) d_{\pi }^{2}  \notag \\
&&+\frac{1}{\pi }\sum_{\nu }\left[ -2\tilde{\Delta}_{\nu }
+\tilde{\Delta}_{\nu }\mathrm{\ln }\left( \frac{\lambda _{\nu }^{2}+\tilde{\Delta}_{\nu
}^{2}}{D^{2}}\right) +2\lambda _{\nu }\mathrm{arctan}
\left( \frac{\tilde{\Delta}_{\nu }}{\lambda _{\nu }}\right) \right] .
\label{ene}
\end{eqnarray}
\end{widetext}

Minimizing Eq. (\ref{ene}) with respect to the Lagrange multipliers one
obtains

\begin{eqnarray}
\lambda _{z} &=&\frac{\tilde{\Delta}_{z}}{\mathrm{\tan }
\left[ \frac{\pi }{2}(1-\sum_{\pi }s_{\pi }^{2})\right] },  \notag \\
\lambda _{\pi } &=&\frac{\tilde{\Delta}_{\pi }}{\mathrm{\tan }
\left[ \frac{\pi }{2}(2s_{\pi }^{2}+3d_{\pi }^{2})\right] }.  \label{lam}
\end{eqnarray}

The derivatives with respect to $s_{\pi }$ and $d_{\pi }$ are lengthy and we
do not reproduce them here. Some simplification is achieved noting that 

\begin{equation}
\frac{\partial \Delta E}{\partial \tilde{\Delta}_{\nu }}=\frac{1}{\pi }
\mathrm{\ln }\left( \frac{\lambda _{\nu }^{2}+\tilde{\Delta}_{\nu }^{2}}
{D^{2}}\right) .  \label{derdel}
\end{equation}
Equating to zero these derivatives and using Eqs. (\ref{lam}) one obtains a
system of four equation with four unknowns $s_{\pi }$ and $d_{\pi }$. \ For
positive splitting $S$, the results for $s_{d}$ and particularly $d_{b}$
indicate if the band of symmetry $b$ is important or not. Furthermore,
according to Eqs. (\ref{si}), (\ref{do}), (\ref{onu}), (\ref{gnu}), 
(\ref{delef}), the low energy part of the spectral density of the molecular
states becomes

\begin{equation}
\rho _{\text{mol}}^{\nu }(\omega )=\frac{\tilde{\Delta}_{\nu }^{2}}{\Delta
_{\nu }\left[ \left( \omega -\lambda _{\nu }\right) ^{2}+\tilde{\Delta}_{\nu
}^{2}\right] },  \label{rhomol}
\end{equation}
which is a Lorentzian centered at $\lambda _{\nu }$ of half width 
$\tilde{\Delta}_{\nu }$ and  weight $\tilde{\Delta}_{\nu }/\Delta _{\nu }$ (the rest
of the spectral weight is at high energies and is not captured by the SBMFA).

\section{Numerical results}

\label{res} In this section we present our results for the solution of the
system of four equations with four unknowns described in the previous
Section. We take the hole Fermi energy $\epsilon_F=0$ as the origin of energies.
Following a previous study for degenerate $\pi $ orbitals \cite{joaq}, 
we take in eV $D=10$, $E_{z}=1$, $E_{a}=2$, and $E_{d}=-2$.
The parameters are justified as follows.
From the diagonalization of the ground-state localized configuration with realistic Coulomb interaction (see for example Ref. \onlinecite{co})
restricted to the three relevant orbitals, it has been established that $E_{a}-E_{z} \sim1 \pm 0.3$ eV in order for the ground state to be a triplet with one hole in the $z^2$ orbital and one in the $\pi$ orbital \cite{joaq}.
A shift in all energies by the same amount is irrelevant. 
In order for the configuration with two holes to be the ground state of
$H_{\mathrm{mol}}$ (isolated molecule) one should have $E_{a}+E_d < E_{z} + \epsilon_F$.
$E_{d}$ was chosen arbitrarily to result in a difference of 1 eV.
However a change in $E_{d}$ and $D$ does not affect the low-energy physics 
(the Kondo effect in particular) if the hoppings are also changed to result in the same
Kondo temperatures.
In order that the result
gives Kondo temperature for the $z^{2}$ channel of the order of the reported
one $\tilde{\Delta}_{z}\sim 20$ meV, we have taken $\Delta _{z}=1.2$ eV.
Similarly in order to obtain  $\tilde{\Delta}_{a}$ 
two orders of magnitude smaller than $\tilde{\Delta}_{z}$, we take 
$\Delta _{\pi }=0.35$ eV as a basis for our study. 

We find that for positive splitting $S$, there is always a local minimum of
the energy for $s_{b}=d_{b}=0$, indicating a 2-channel situation. As
expected, this local minimum is not the global minimum for small $S$. In
this case the weight of the singly occupied states for both $\pi $ channels
are similar ($s_{b}^{2}\sim s_{a}^{2}$) and the same happens for the doubly
occupied sites ($d_{b}^{2}\sim d_{a}^{2}$). We expected that for large $S$
there would be a global minimum for small non-zero $s_{b}$,$d_{b}$, but this
is not the case. There is an abrupt jump in the position of the global
minimum between two local minima, one with $s_{b}=d_{b}=0$ and the other one
with $s_{b}^{2}\sim s_{a}^{2}$ and $d_{b}^{2}\sim d_{a}^{2}$.

\begin{figure}[h]
\begin{center}
\includegraphics[width=7cm]{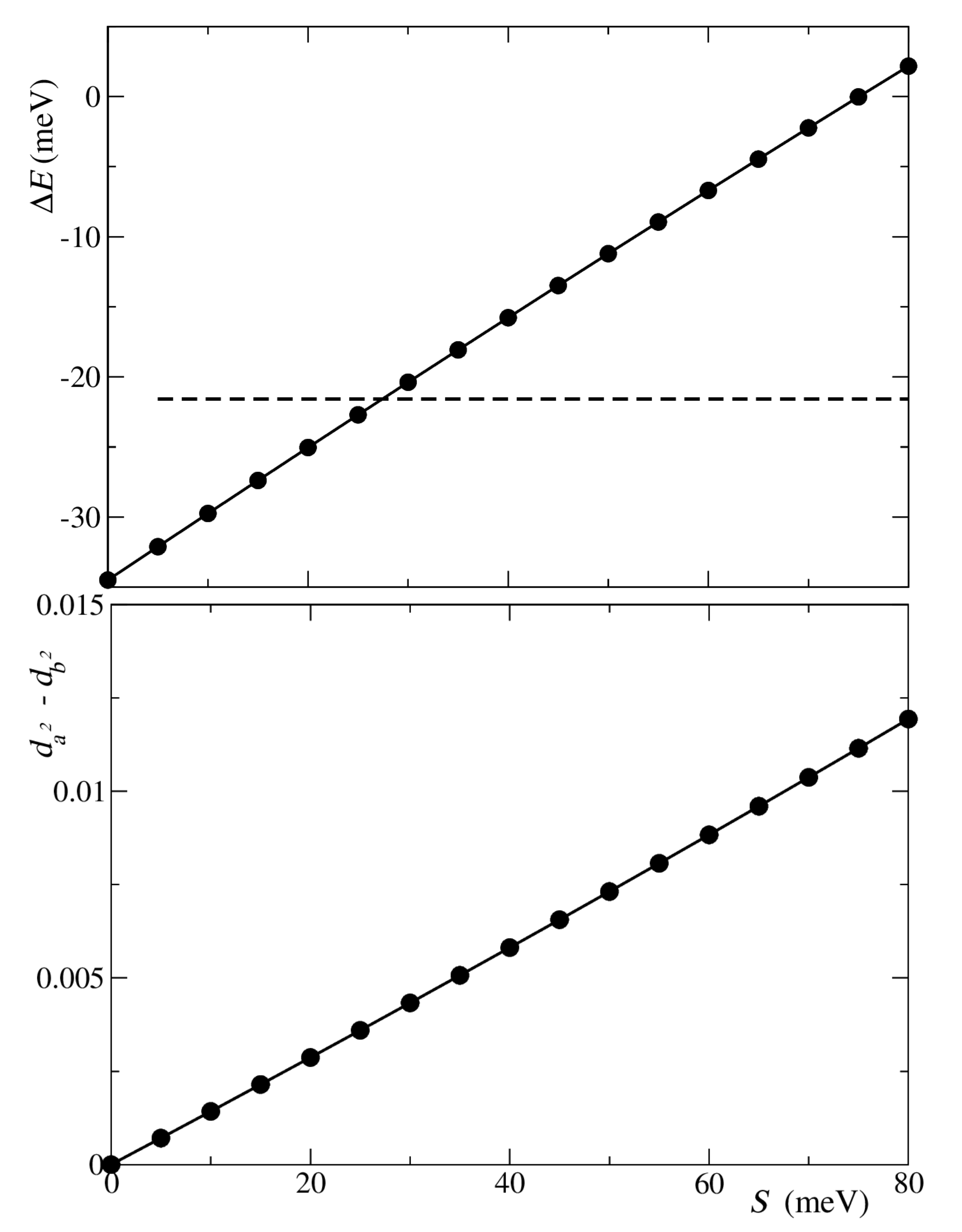}
\caption{Energy (top) and orbital polarization (bottom) for the 3-channel solution as a function of the splitting $S$. The dashed line indicates the energy of the 2-channel solution.}
\label{fig}
\end{center}
\end{figure}

In Fig. \ref{fig} we represent the energy  
and  the difference $P=d_{a}^{2}- d_{b}^{2}$ 
for the 3-channel solution as a function of the splitting $S$.
For the 2-channel solution, the energy is constant at the value $\Delta E=-21.5745$ eV.
Similarly in this solution  $d_{a}^{2}=0.317$ (slightly below 1/3 as expected, 
since the weight is shared by all three spin projections of the triplet) and $d_{b}^2=0$. 
For both solutions in general $d_{a}^{2}+d_{b}^{2} \sim 1/3$,  
and $s_{z}^{2}$ and $s_{\pi}^{2}$ are very small. 
Therefore $3P$, which is the total orbital polarization
of the ground state, is the more relevant bosonic variable.

As observed in the figure, the orbital polarization of the 3-channel solution 
increases almost linearly (the curvature is small and positive) with
$S$, but its magnitude is very small (less than 5\% 
of the maximum value in the range studied) and as a consequence 
the energy increases fast with $S$ (almost linearly with a negative second derivative).
For the parameters chosen, the ground state of the isolated molecule $H_{\mathrm{mol}}$ 
is $E_a+E_d=0$. 
Therefore, a positive $\Delta E$ (as we obtain for the 3-channel solution for large $S$)
would indicate that including the mixing terms $H_{\mathrm{mix}}$ 
of the molecule with the conduction electron 
{\em increases} the energy of the system. This is nonphysical and points out
the instability of the 3-channel solution for large $S$.
Actually, at the critical splitting $S_c=27.07$ meV there is a transition
to the 2-channel solution which becomes that of lower energy for $S>S_c$.

\begin{table}[h]
\caption{\label{tab1} Results for different observables at the 
transition between both phases for $\Delta _{\pi }=0.35$ eV. 
The 2- (3-)channel results are 
above (below) the horizontal line in the middle. }
\begin{ruledtabular}
\begin{tabular}{llll}
  & $z$ & $a$ & $b$   \\ \hline
$s^2_\nu$  & $6.87 \times 10^{-4}$ & 0.0232     & 0  \\
$d^2_\pi$    & - & 0.317 & 0  \\
$\tilde{\Delta}_{\nu }$ (meV)  & 25.7 & 0.222 & 0  \\
$\lambda_{\nu }$ (meV)         & 1.88 & $4.80 \times 10^{-4}$ & 0  \\ \hline
$s^2_\nu$  & 0.0235 & 0.00727     & 0.00691  \\
$d^2_\pi$    & - & 0.156 & 0.152  \\
$\tilde{\Delta}_{\nu }$ (meV) & 15.3 & 6.53 & 6.36  \\
$\lambda_{\nu }$ (meV)   & 0.681 & 6.89 & 6.98  \\ 
\end{tabular}
\end{ruledtabular}
\end{table}

The results for the different quantities at the transition are indicated in
Table \ref{tab1}. In the 2-channel solution, the half width of the resonance
for the molecular states of   $z^{2}$ symmetry (identified with the
respective Kondo temperature) is $\tilde{\Delta}_{z}=25.7$ meV, somewhat
larger than reported previously $\tilde{\Delta}_{z}\sim 20$ meV \cite{mina}.
However, comparison with theory suggest that  $\tilde{\Delta}_{z}>20$ meV 
\cite{nl}. The position of this peak ($\lambda _{z}=1.88$ meV) is
practically at the Fermi energy. The half width of the peak for the
molecular $a$ state is $\tilde{\Delta}_{a}=0.222$ meV and it lies at the
Fermi energy. The molecular $b$ sate is absent at low energies in this
solution. Note that the weight of the singly occupied  $a$ \ states is
related to the valence fluctuations of the $z^{2}$ states and vice versa.
Therefore  $\tilde{\Delta}_{z}>\tilde{\Delta}_{a}$ implies 
$s_{a}^{2}>s_{z}^{2}$ ($s_{a}^{2}=0.023$, $s_{z}^{2}\sim 7\times 10^{-4}$ in
this case).

The 3-channel  solution is markedly different. The $\pi $ ($a$ and $b$)
channels behave as quasi degenerate. The weight of these channels in the
ground state configuration of doubly occupied states is very similar 
($d_{a}^{2}=0.156$, $d_{b}^{2}=0.152$). This fact has an effect of increasing
markedly the Kondo temperature of these channels, as expected for example
when the symmetry of the SU(2) Kondo model is increased to SU(4) \cite{su42,lopes,tera}. We obtain 
$\tilde{\Delta}_{a}=6.53$ meV, $\tilde{\Delta}_{a}=6.36$ meV. The
corresponding peaks are shifted from the Fermi energy (below it in the
electron representation as opposed to the hole one used here) by  
$\lambda_{a}=6.89$ meV, $\lambda _{b}=6.98$ meV. The fact that the position and the
half width of the peaks are of the same order is also expected, for example
from the SU(4) Anderson model \cite{su42,lopes,tera}. The increase of the Kondo energy scale for the $\pi $ channels has the effect of decreasing 
the corresponding scale for the  $z^{2}$ channel. This
competition has been studied before for degenerate $\pi $ channels 
\cite{joaq}. As a consequence the half width of the molecular state with $z^{2}$
symmetry is reduced to $\tilde{\Delta}_{z}=15.3$ meV.  

\begin{table}[h]
\caption{Same as Table \protect\ref{tab1} for 
$\Delta _{\protect\pi }=0.4$ eV.}
\label{tab2}
\begin{ruledtabular}
\begin{tabular}{llll}
  & $z$ & $a$ & $b$   \\ \hline
$s^2_\nu$  & $2.21 \times 10^{-3}$ & 0.0228     & 0  \\
$d^2_\pi$    & - & 0.317 & 0  \\
$\tilde{\Delta}_{\nu }$ (meV)  & 25.2 & 0.815 & 0  \\
$\lambda_{\nu }$ (meV)         & 1.81 & $5.66 \times 10^{-3}$ & 0  \\ \hline
$s^2_\nu$  & 0.0322 & 0.00591     & 0.00544  \\
$d^2_\pi$    & - & 0.155 & 0.149  \\
$\tilde{\Delta}_{\nu }$ (meV) & 12.1 & 10.0 & 9.58  \\
$\lambda_{\nu }$ (meV)   & 0.431 & 10.7 & 10.9  \\ 
\end{tabular}
\end{ruledtabular}
\end{table}

The critical splitting for the transition $S_{c}=27.07$ meV is markedly
smaller to the SOC of Fe $\sim 76$ meV \cite{socfe} In order to see the
sensitivity of the results with $\tilde{\Delta}_{a}$ (which has some
uncertainty) and in particular if $S_{c}$ can be increased substantially, we
have increased the resonant-level width of the $\pi $ states to 
$\Delta_{\pi }=0.4$ eV. The new critical splitting becomes $S_{c}=46.77$ meV. 

The corresponding results for the different quantities at this value of the
splitting are listed in Table \ref{tab2}. The main change in the 2-channel
region is that the Kondo temperature of the $a$ channel is increased by a
factor near 4 to $\tilde{\Delta}_{a}=0.815$ meV, which seems too large for
an agreement with experiment \cite{nl}. The corresponding result for the  
$z^{2}$ channel is only moderately decreased to $\tilde{\Delta}_{z}=25.2$ meV.

The changes in the 3-channel region are moderate and expected. 
$\tilde{\Delta}_{a}$ and $\tilde{\Delta}_{b}$ increase to $\sim 10$ meV, 
$\tilde{\Delta}_{z}$ decreases to 12 meV, and correspondingly $s_{z}^{2}$ decreases and $s_{\pi }^{2}$ decrease.

We have also studied the case $\Delta _{\pi }=0.3$ eV.  \ The critical
splitting decreases to $S_{c}=12.94$ meV. The values of the different
observables at the transition are displayed in Table \ref{tab3} In the
2-channel solution, $\tilde{\Delta}_{a}$ decreases by a factor near 5 with
respect to the case shown in Table \ref{tab1} (for which $\Delta _{\pi }=0.35$ eV). 
$\tilde{\Delta}_{z}$ increases in 1\%. In the 3-channel solution, 
$\tilde{\Delta}_{z}=19$ meV is more similar to the value of the 2-channel solution
and  $\tilde{\Delta}_{\pi }\sim 3.5$ meV with near 1\% difference between 
$\tilde{\Delta}_{a}$ and $\tilde{\Delta}_{b}$ (they tend to be equal due to
the decrease of the splitting).

\begin{table}[h]
\caption{Same as Table \protect\ref{tab1} for $\Delta _{\protect\pi }=0.3$
eV. }
\label{tab3}
\begin{ruledtabular}
\begin{tabular}{llll}
  & $z$ & $a$ & $b$   \\ \hline
$s^2_\nu$  & $1.37 \times 10^{-4}$ & 0.0233     & 0  \\
$d^2_\pi$    & - & 0.318 & 0  \\
$\tilde{\Delta}_{\nu }$ (meV)  & 25.9 & 0.038 & 0  \\
$\lambda_{\nu }$ (meV)         & 1.90 & $1.63 \times 10^{-5}$ & 0  \\ \hline
$s^2_\nu$  & 0.0145 & 0.00882     & 0.00861  \\
$d^2_\pi$    & - & 0.157 & 0.155  \\
$\tilde{\Delta}_{\nu }$ (meV) & 19.0 & 3.54 & 3.50  \\
$\lambda_{\nu }$ (meV)   & 1.04 & 3.67 & 3.70  \\ 
\end{tabular}
\end{ruledtabular}
\end{table}

\section{Summary and discussion}

\label{sum}

We have studied a generalized Anderson model in which two triplets are
hybridized with three higher energy doublets, with a variable splitting $S$
between both triplets assumed to be the same as that between two doublets.
The model contains three hybridizing channels and  has been proposed to
describe the low-energy  physics of an isolated iron phthalocyanine
molecule deposited on the Au(111) surface. The triplets contain
one hole in the Fe 3d orbital with $z^{2}$ symmetry an another one in one of
the 3d $\pi $ orbitals.
The split $\pi $ orbitals are orthogonal linear combinations of $xz$ and $yz$ 
orbitals. If the origin of the splitting is spin-orbit coupling, 
these combinatios are given by Eqs. (\ref{states}).
The doublets have one hole in
any of the three molecular orbitals. The different channels correspond to
the three different symmetries.

Clearly, for large $S$ one of the $\pi $ channels can be neglected at low
energies and the model can be reduced to a 2-channel type, justifying
previous calculations in which several experiments were explained on the
basis of a 2-channel spin-1 Kondo model with easy plane anisotropy \cite{nl}.
These calculations are particularly interested because they imply that the
system is a topologically non-trivial  \textit{non-Landau} Fermi liquid. 
We have not included the anisotropy here to avoid entering the topologically
non-trivial phase which cannot be described by the approach.

The question we have addressed is how large should $S$ be to justify this
further low-energy reduction to a 2-channel model.
Our results using a slave-boson mean-field approximation predict an abrupt
transition from a 3-channel to a 2-channel regime with increasing splitting.
While the first-order nature of the transition is probably an artifact of the
mean-field approximation, we believe that the resulting critical splitting 
$S_{c}$ has semiquantitative validity. For the parameters which best
correspond to experimental observations we obtain $S_{c}\sim 27$ meV. This
is substantially smaller than the spin-orbit coupling of Fe  $\sim 76$
meV \cite{socfe}. Therefore we expect that in fact the 2-channel model is
justified, and the relevant $\pi $ channel corresponds to the $a$ states
described in Eq. (\ref{states}). The effective SOC might be reduced 
by a few percent due to the admixture of some amount of N $p$ orbitals 
in the molecular states \cite{lobos}, but this does not 
affect our conclusions.

The reduction of the model from 3-channel to 2-channel due to spin-orbit 
coupling has other consequences in the comparison to experiment. 
The states of the ground 
state configuration with spin projection $S_{z}=\pm 1$ have also angular
momentum projection $L_{z}=\pm 1$ [see Eqs. (\ref{states}) and below them]. Therefore, the effect of a magnetic field 
$B_{z}$ in the $z$ direction (perpendicular to the Au(111) surface as applied
experimentally \cite{yang}), with a correction term  $(2S_{z}+L_{z})B_{z}$, is
3/2 larger than the case in which only the spin is considered. In addition
the effect of magnetic field perpendicular to $z$, involving spin flips
induces mixing with excited $b$ states and is smaller than for the case without
splitting.

\section*{Acknowledgments}
We thank J. Lorenzana and J. Fern\'andez for useful discussions.
We acknowledge financial support provided by PICT 2017-2726 and PICT 2018-01546 of the ANPCyT, Argentina.

\end{document}